\documentstyle[12pt, epsf, bookman]{article}
\def\singlespace {\smallskipamount=3.75pt plus1pt minus1pt
                  \medskipamount=7.5pt plus2pt minus2pt
                  \bigskipamount=15pt plus4pt minus4pt
                  \normalbaselineskip=15pt plus0pt minus0pt
                  \normallineskip=1pt
                  \normallineskiplimit=0pt
                  \jot=3.75pt
                  {\def\smallskip {\vskip\smallskipamount}}
                  {\def\medskip   {\vskip\medskipamount}}
                  {\def\bigskip   {\vskip\bigskipamount}}
                  {\setbox\strutbox=\hbox{\vrule
                    height10.5pt depth4.5pt width 0pt}}
                  \parskip 7.5pt
                  \normalbaselines}
\def\middlespace {\smallskipamount=5.825pt plus1.5pt minus1.5pt
                  \medskipamount=11.25pt plus3pt minus3pt
                  \bigskipamount=22.5pt plus6pt minus6pt
                  \normalbaselineskip=22.5pt plus0pt minus0pt
                  \normallineskip=1pt
                  \normallineskiplimit=0pt
                  \jot=5.825pt
                  {\def\smallskip {\vskip\smallskipamount}}
                  {\def\medskip   {\vskip\medskipamount}}
                  {\def\bigskip   {\vskip\bigskipamount}}
                  {\setbox\strutbox=\hbox{\vrule
                    height15.75pt depth6.75pt width 0pt}}
                  \parskip 7.25pt
                  \normalbaselines}
\def\dblspc {\smallskipamount=7.5pt plus2pt minus2pt
                  \medskipamount=15pt plus4pt minus4pt
                  \bigskipamount=30pt plus8pt minus8pt
                  \normalbaselineskip=30pt plus0pt minus0pt
                  \normallineskip=2pt
                  \normallineskiplimit=0pt
                  \jot=7.5pt
                  {\def\smallskip {\vskip\smallskipamount}}
                  {\def\medskip   {\vskip\medskipamount}}
                  {\def\bigskip   {\vskip\bigskipamount}}
                  {\setbox\strutbox=\hbox{\vrule
                    height21.0pt depth9.0pt width 0pt}}
                  \parskip 15.0pt
                  \normalbaselines}
\def\nb{\nabla }

\def\be{\begin{equation}}

\def\j-{\J_-}
\def\al{\alpha}

\def\eps{\epsilon}
\def\gm{\gamma}

\def\ee{\end{equation}}

\def\bearr{\begin{eqnarray}}
\def\bearrs{\begin{eqnarray*}}
\def\eearr{\end{eqnarray}}
\def\eearrs{\end{eqnarray*}}
\def\barr{\begin{array}}
\def\earr{\end{array}}
\def\p{\partial}

\def\th{\theta}

\def\o{\omega}

\def\non\non{\nonumber}
\def\nn8{\nonumber\\[15pt]}
\def\l{\left}
\def\r{\right}

\def\f{\frac}

\begin{document}
\thispagestyle{empty}
 \middlespace

\begin{center}
\vspace*{2.5in}
 {\huge Inertial Forces - \'a la Newton\\[8pt]
 in General
Relativity}\footnote{Lecture delivered at the meeting on `Black Hole Astrophysics', held in March
2001, at Calcutta}\\[30pt]

A.R. Prasanna\\ Physical Research Laboratory\\ Navrangpura,
Ahmedabad 380 009\\ India\\[40pt]
\end{center}
\newpage
\section{Introduction}Einstein, generalising the special theory of
relativity to include gravitational force gave the beautiful
theory of general relativity wherein the space-time geometry took
the center stage with its curvature personifying   gravitation.
As a covariant theory General relativity is completely independent of
observers and their state of motion. Thus it is always interesting
to look for adapting systems of coordinates wherein one can find
some new interpretation that could help in understanding the
physical behaviour of a system better. In fact it is well known that
in the full 4 dim. space-time theory General Ralativity dispensed
away with the Newtonian notion of inertial forces. However in
1988, Abramowicz Carter and Lasota showed that if one introduces
the so called optical geometry with a 3+1 splitting  of space-time
using a conformal slicing, then one can reintroduce the notion of
Newtonian forces, which can be useful in clarifying the origin of
certain GTR effects that were known in the 4-dim. theory. For a
free particle only under the influence of gravity the equations of
motion are given by the geodesics of the given spacetime manifold
describing the gravitational field. Much as one admires the
language of geometry the question is, does one get all the
information inherent in the system within this framework or could
there be something overlooked or missed in the interpretation.
Further, the language of forces has indeed been very useful in
describing the other interactions and thus it could be useful to
bring it back into the realms of  General Relativity, which might
focus the non-Newtonian features inherent in the geometry more
explicitly. This approach would particularly be of help in studying
physical aspects very close to ultra compact objects and
blackholes, a region which is never available for the weak field
Newtonian physics.
\section{Formalism} As the idea is to bring in
Newtonian language into a geometric theory of spacetime, one needs
to slice the 4-space into a (3 space + time) structure and look at
the features on the absolute 3 space so obtained. In fact, such a
3+1 split of spacetime is nothing new in general relativity, as,
long ago Arnowitt, Deser and Misner (1962) introduced such a
scheme while looking for a method to give a Hamiltonian descriptin
of general relativity. As Misner, Throne and Wheeler (1972)
mention, ``{\it The slicing of spacetime into a one parameter
family of space-like hypersurfaces is called for, not only by the
analysis of the dynamics along the way, but also by the boundary
conditions as they pose themselves in any action principle of the
form - give the 3-geometries on the two faces of a sandwich of
spacetime and adjust the 4-geometry in between to extremize the
action}".\\

In fact, such a procedure of studying the dynamics effectively
paved the way for setting up numerical methods to study
evolution equations of the Cauchy data given on an initial hyper
surface, and thus became a standard procedure for numerical
relativity (Seidel (1996) and York (1979)).\\

Instead of a fully dynamical system, suppose one has a
stationary system wherein a time-like Killing vector exists then
one can get a lower dimensional quotient space through an
isometry group action and one can study certain dynamical
features within a given geometrical background.  Abramowicz,
Carter and Lasota (1988, hereafter referred to as ACL) used such
a prescription with a conformal rescaling factor and showed that
one can indeed obtain a 3+1 splitting wherein the 3-space is the
quotient space obtained from the action of the time-like Killing
vector and the metric conformal to the spatial geometry of the
original four-space.  As they realised the most significant
feature of a conformal reslicing was that the normally
geometrical geodesic equation for a test particle would separate
into language of Newtonian forces wherein one can directly
interpret terms as graviational, centrifugal and Coriolis
accelerations.\\

Abramowicz, Nurowski and Wex (1993, hereafter referred to as
ANW) later gave a covariant approach to this formulation which
does not depend upon any particular symmetry and is as follows:\\

In the given spacetime manifold $M$ with the metric
\be
ds^2 = g_{ij} dx^idx^j
\ee
introduce a congruence of world lines which is globally
orthogonal to $t$ = const., hypersurface which ensures that the
vorticity of the congruence to be zero.  In fact, Bardeen (1972)
adopted such a congruence in axisymmetric, stationary spacetimes
defining what are called locally non-rotating observers or zero
angular momenturm observers (ZAMO).  The advantage of having
such a congruence is that these local observers `rotate with the
geometry' and the connecting vectors between two such observers
with adjacent trajectories do not precess with respect to
Fermi-Walker transport.  Denoting such a vector field by $n^i$
$\l( n_i n^i =-1 \mbox{ time-like} \r)$ it can be verified that
the corresponding four-acceleration is proportional to the
gradient of a scalar potential.
\be
n^k \nb_k n_i = \nb_i \phi \; ; \; \; \; n^i \nb_i \phi = 0 \ee
Though the vector field $n^i$ is not uniquely determined by (2),
locally each particular choice of $n^i$ uniquely defines a
foliation of the spacetime into slices each of which represents
space at a particular instant of time, whose geometry is given by
\be
h_{ik} = g_{ik} + n_i n_k \; ; \; h^i_{\; k} = \delta^i_{\; k} +
n^i n_k \ee (2) also ensures that the special observers $n^i$ see
no change in the potential as their proper time passes by and thus
have fixed positions that help them in distinguishing between
different `inertial forces'.\\

Consider a particle of rest mass $m_o$ and four-velocity $U^i$,
which can be expressed as
\be
U^i = \gm \l( n^i + v \tau^i \r) \ee wherein $\tau^i$ is the unit
tangent vector (space-like) orthogonal to $n^i$ and parallel to
the 3-velocity $v$ of the particle in the 3-space (Lorentz speed)
and $\gm$ the Lorentz factor $\l( = \f{1}{\sqrt{1-v^2}}\r)$. The
four-acceleration of the particle $a^i$ may now be obtained
through direct computation (Abramowicz, 1993)
\be
\barr{lll} a_k&=& u^i \nb_i u_k = - \gm^2 \nb_k \phi +\gm^2 v \l(
n^i \nb_i \tau_k + \tau^i \nb_i n_k \r)\\[8pt] &&+ \gm^2 v^2
\tau^i \nb_i \tau_k + (v \gm )^. \tau_k + \dot{\gm} n_k \earr \ee
Let us consider the motion of the particle in a circular orbit in
a general stationary axisymmetric spacetime.  From the given
symmetries, there exist two Killing vectors $\eta^i$ the time-like
having open trajectories, and $\zeta^i$ the space-like with closed
trajectories.  If the particle has a constant angular velocity
$\Omega$ as measured by the stationary observer at infinity then
its four-velocity $u^i$ may be expressed as
\be
u^i = A \l( \eta^i + \Omega \zeta^i \r)
\ee
$A$ being the redshift factor
\be
A^2 = \l[ < \eta \eta > + 2 \Omega < \eta \zeta > + \Omega^2 <
\zeta \zeta > \r]^{-1}
\ee
In terms of the Killing vectors one can express $n^i$ and $\phi$
consistently as
\be
n^i = e^\phi \l( \eta^i + \o \zeta^i \r) \; , \; \; \; \o = - <
\eta , \zeta >/<\zeta , \zeta >
\ee
and
\be
\phi = - \f{1}{2} \ell n \l[ - < \eta , \eta > - 2 \o < \zeta ,
\eta > - \o^2 < \zeta , \zeta > \r] \ee From (4) and (6) one can
evalutate the particle speed $V$ to be:
\be
V\tau^i = e^\phi ( \Omega - \omega ) \zeta^i
\ee
Using now the ACL approach of conformal rescaling of the
3-metric, one can define the projected metric and the vectors
\be
\tilde{h}_{ij} = e^{2\phi} \; h_{ij} \; , \; \; \tilde{\tau}^i =
e^\phi \tau^i \ee such that the acceleration may now be written as
\be
a_k: = - \nb_k \phi + \gm^2 V \l( n^i \nb_i \tau_k + \tau^i \nb_i
n_k \r) + (\gm V)^2 \tilde{\tau}^i \tilde{\nb}_i \tilde{\tau}_k
\ee as the last two terms in (5) become zero for a particle with
constant speed, and conserved energy. $\tilde{\nb}$ in (12) refers
to the covariant derivative with respect to the metric $\tilde{h}_{ij}$.
As may be seen, the acceleration is made up of three distinct
terms, (i) gradient of a scalar potential, (ii) a term
proportional to $V$, and (iii) one proportional to $V^2$. The
first and the third terms may be immediately recognised as the
gravitational and centrifugal accelerations.  Further, as $n^i$
and $\tau^i$ are parallel to the Killing vector $\eta^i$ and
$\zeta^i$ respectively from the equation for Lie derivative, one
has
\be
L_n \tau \equiv n^i \nb_i \tau_k + \tau_i \nb_k n^i = 0 \ee Using
this alongwith the fact that $n^i$ and $\tau^i$ are orthogonal,
the second term of (12) may be written as
\be
\gm^2 \l[ n^i \l( \nb_i \tau_k - \nb_k \tau_i \r) \r] \ee which
represents the Lense-Thirring effect of the inertial drag and thus
identified as the generalisation of the Coriolis acceleration.\\

If the general axisymmetric and stationary spacetime is
represented by the metric
\be
ds^2 = g_{tt} dt^2 + 2 g_{t\phi} dt d\phi + g_{\phi\phi} d\phi^2
+ g_{r} dr^2 + g_{\th\th} d\th^2
\ee
with $g_{ij}$s being functins of $r$ and $\th$ only then the
total force acting on a particle in cirular orbit with constant
speed $\Omega$ may be expressed as (Prasanna (1997))
\be
F_i: = (Gr)_i + (Co)_i + (Cf)_i
\ee
wherein
\be
\barr{lll} (Gr)_i:&=& - \nb_i\phi = \f{1}{2} \p_i \l\{ \ell n \l[
\l( g^2_{t\phi} - g_{tt} g_{\phi\phi} \r) / g_{\phi\phi} \r]
\r\}\\[8pt] (Co)_i:&=& \gm^2 Vn^k \l( \nb_k \tau_i - \nb_i \tau_k
\r)\\[8pt] &=& - a^2 (\Omega - \o ) \sqrt{g_{\phi\phi}} \l\{ \p_i
\l( \f{g_{t\phi}}{\sqrt{g_{\phi\phi}}}\r) + \o \p_i
\sqrt{g_{\phi\phi}} \r\} \earr \ee and
\be
\barr{lll}
(Cf)_- &=& (\gm V)^2 \tilde{\tau}^k \tilde{\nb}_k \tilde{\tau}_i\\[8pt]
&=& - A^2 (\Omega - \o )^2 \f{g_{\phi\phi}}{2} \pi_i \l\{ \ell n
\l[ \left.  g^2_{\phi\phi} \right| \l( g^2_{t\phi} -
g_{tt}g_{\phi\phi} \r) \r] \r\}
\earr
\ee
with
\[
\Phi = - \f{1}{2} \ell \l[ - g_{tt} - 2\o g_{t\phi}  - \o^2
g_{\phi\phi}\r] \; ; \; \o = - g_{tq} / g_{\phi\phi}
\]
and
\be
A^2 = - \l[ g_{tt} + 2\Omega g_{t\phi} + \Omega^2 g_{\phi \phi} \r]^{-1}
\ee
In order to understand this splitting of the total force, let us
consider the simple case of static spacetimes (Abramowicz and
Prasanna (1990)). Then, by definition the Coriolis term would be
absent, and the force acting on a test particle of mass $m_o$
and 3-momentum $p^\al$ is given by (ACL)
\be
m_o\tilde{f}_\al = \tilde{p}^\mu \tilde{\nb}_\mu \tilde{p}_\al +
\f{m^2_o}{2} \tilde{\nb}_\al \phi \ee wherein $\tilde{p}^\mu = m_o
V \gm \tilde{\tau}^\mu$ and thus consistent with what one gets
from (15) in the projected 3-space.  It is now clear that in the
static space time, the trajectories of photons (rest mass zero
particle) are given by the curves
\be
\tilde{\tau}^\mu \tilde{\nb}_\mu \tilde{\tau}_\al = 0
\ee
which by definition are geodesics of this 3-space.  Thus, one
finds that the `null trajctories' of the 4-space project onto
the geodesics of the quotient space obtained through conformal
slicing and indicates that particles on these trajectories do
not experience any `force' in the Newtonian sense.\\

It is for this reason that ACL called this slicing as optical
reference geometry meaning that the null lines project onto
straight lines in the Euclidean sense. However, as was seen
later this is true only for static spacetimes, as rotation
would influence the photons differently for prograde and
retrograde motion.  In static spacetimes, one finds
that particles acted on by forces other than gravity would
deviate from the geodesics of the quotient 3-space obtained by
conformal slicing, as these geodesics behave like straight lines
of Newtonian geometry and thus follow Newtonian laws of motion.\\

The most important aspect of the frmalism is that while the
particle kinematics is expressed in Newtonian language, the
general relativistic effects are all present and thus help in
understanding results which were earlier known in general
relativistic analysis but were hidden in the geometry and not
accessible for visualisation.\\
\section{Specific Applications}
We start from the simplest application of the methodology
outlined above to study the particle kinematics in the static
spacetimes, taking the Schwarzschild geometry as the first
example (Abramowicz and Prasanna, 1990; hereafter referred to as
AP).\\

The metric as expressed in the usual coordinates,
\be
ds^2 = - \l( 1 - \f{2m}{r} \r) dt^2 + \l( 1 - \f{2m}{r} \r)^{-1}
dr^2 + r^2 \l( d\th^2 + \sin^2\th d\phi^2 \r)
\ee
would yield for the gravitational and centrifugal accelerations,
acting on a test particle in ciruclar orbit, the expressions:
\be
(Gr)_r = \f{m}{r^2} \l( 1 - \f{2m}{r} \r)^{-1}
\ee
and
\be
(Cf)_r = - \Omega^2 r \l( 1 - \f{2m}{r}\r)^{-1} \l( 1 - \f{2m}{r}
- \Omega^2 r^2 \r)^{-1} \l( 1 - \f{3m}{r} \r) \ee From (24) it is
clear that while $(Gr)$ and $(Cf)$ are in opposite directions upto
$r = 3m$ from infinity for $r < 3m$ they both have the same
direction. As gravitational acceleration is always undirectional,
it is clear that the centrifugal acceleration reverses its sign at
$r = 3m$. In fact, if one looks at the photon effective potential
in the Schwarzschild geometry, one finds that $r = 3m$ is the
location of the maximum and thus corresponds to unstable circular
orbit. What has been noticed now is that this null line is the
straight line path for the photon in the quotient space and thus
corresponds to a location at which the centrifugal acceleration is
zero, and on either sides the centrifugal force acts in opposite
directions.\\

As shown in AP, this feature has important kinematical
implications in the study of accretion flows near ultra compact
objects (black holes) like the Rayleigh ceiterion for stability
of flow turns out to be
\be
\f{2m \l( r^3 - 2 mr^2 \r)}{\l( r^3 - \ell^2 r + 2m \ell^2
\r)^2} \l( 1 - \f{3m}{r} \r) \f{d\ell^2}{dr} > 0
\ee
This means $\f{d\ell^2}{dr} > 0$ for $r > 3m$ and $< 0$ for $r <
3m$, indicating that for $r < 3m$ the angular momentum has to be
advected inwards for stability.  In fact, this result clearly
explained the findings of Anderson and Lemos (1988), who had
obtained inward advection of angular momentum very close to
black holes.\\

Another important implication of centrifugal reversal is borne
out in the evaluation of ellipticity of slowly rotating fluid
configuration represented by a sequence of quasi-stationary
solutions, with decreasing radii, keeping the mass and angular
momentum conserved.  Chandrasekhar and Miller (1974) had
considered this problem and found that the ellipticity instead
of increasing continuosly, reached a maximum, a result which
they had attributed to frame dragging of rotatintg systems.
However, after the discovery of centrifugal reversal, Abramowicz
and Miller (1990) analysed the equilibrium configuration in
general relativity using the `inertial forces' approach and
found that the equilibrium demands
\be
\barr{lll}
\f{GM}{R^2}& =& R\Omega^2_k \l( 1 -\f{3m}{R} \r) \l( 1 - \f{2m}{R}
- R^2 \Omega^2_k \r)^{-1}\\[8pt]
\Omega^2_k&=& \f{GM}{R^3}
\earr
\ee
which to the lowest order in $\Omega$ gave the ellipticity
function to be
\be
\eps (R) = \f{125}{32} \l[ 1 - \f{3}{2R} \r] \l( \f{1}{R} \r) \ee
exhibiting a maximum at $R=3$.\\

Though they had obtained the ellipticity maximum, the lacuna in
their approach was that they had used only the Schwarzschild
exterior geometry for evaluating the inertial forces.  Prasanna
and coworkers (A. Gupta, S. Iyer and A.R. Prasanna (1996))
reconsidered this problem starting from the general conservation
laws expressed in the 3+1 formalism and using the Hartle-Thorne
approximation solution for the interior of a slowly rotating
body. Defining the ellipticity through force balance equations
at the pole and equator
\be
e^2 = 1 \f{\l( F_{cf} - F_{ge}\r)}{F_{gp}}\; , \; \; \eps \approx
\f{e^2}{e} \ee ($F_{ge}$ is the gravitational force at the equator
and $F_{gp}$ at the poles) they found the maximum to occur at $r =
5.4m$, which is much closer to Chandrasekhar-Miller result of
maximum occurring at $r =5m$.\\

Prasanna (1991) had considered another static spacetime viz.
Ernst spacetime which represents the external field of a
blackhole immersed in a uniform magnetic field. The photon
effective potential in this metric has two extrema, the maxima
corresponding to the unstable orbit very close to $r = 3m$,
while a minima located far away, governed by the strength of the
magnetic field and yielding a stable photon orbit.  As
centrifugal force would go to zero at both these locations, it
appears that the Newtonian orbits are possible only within the
region bounded by these two points of extrema.\\

We next consider the situation in the Kerr geometry which is
actually supposed to be the spacetime exterior to a rotating
blackhole represented by the metric
\be
\barr{lll}
ds^2&=& \l( 1 - \f{2mr}{\Sigma} \r) dt^2 - \f{4amr}{\Sigma}
\sin^2\th dt d\phi + \f{\Sigma}{\Delta} dr^2 + \Sigma d\th^2\\[8pt]
&&+ \f{B}{\sigma} \sin^2\th d\phi^2
\earr
\ee
with
\[
\barr{lll}
\Delta&=& r^2 - 2mr + a^2\\[8pt]
\Sigma&=& r^2 + a^2 \cos^2 \th\\[8pt]
B &=& \l( r^2 + a^2 \r) - \Delta a^2 \sin^2\th
\earr
\]
Using the formluae presented in (16)-(18), one can get (Prasanna
(1997))
\be
(GR)_r = \f{m \l( a^2 \Delta + r^4 + a^2 r^2 - 2ma^2r
\r)}{r\Delta \l( r^3 + a^2 r + 2ma^2 \r)}\\
\ee

\be
(Cf)_r = \f{ - (\Omega - \o )^2 \l[ r^5 - 3m r^4 + a^2 \l( r^3 -
3mr^2 + 6m^2r - 2ma^2 \r) \r]}{r^2 \Delta \l[ 1 - \Omega^2 \l( r^2 + a^2
\r) - \l( \f{2m}{r} \r) \l( 1 - \Omega a \r)^2 \r]}\\
\ee

\be
(Co)_r = \f{2ma (\Omega - \o ) \l( 3r^2+a^2 \r)}{r \l( r^3 + a^2r +
2ma^2 \r) \l[ 1 - \Omega^2 \l( r^2 + a^2 \r) - \l( \f{2m}{r} \r)
( 1 - \Omega a)^2\r]}\\
\ee

One can immediately see the difference
in the nature of centrifugal and Coriolis forces,
whereas the
Coriolis depends on the coupling of the angular
momentum of the central source with that of the particle $a(\Omega
- \o )$, the centrifugal can go to zero at different
locations solely depending upon `$a$' due to the zeros of the quintic
equation
\be
r^5 - 3mr^4 + a^2 \l( r^3 - 3mr^2 + 6m^2 r - 2ma^2 \r) = 0 \ee

\begin{center}
\epsfxsize 4in \epsfysize 4in \epsfbox{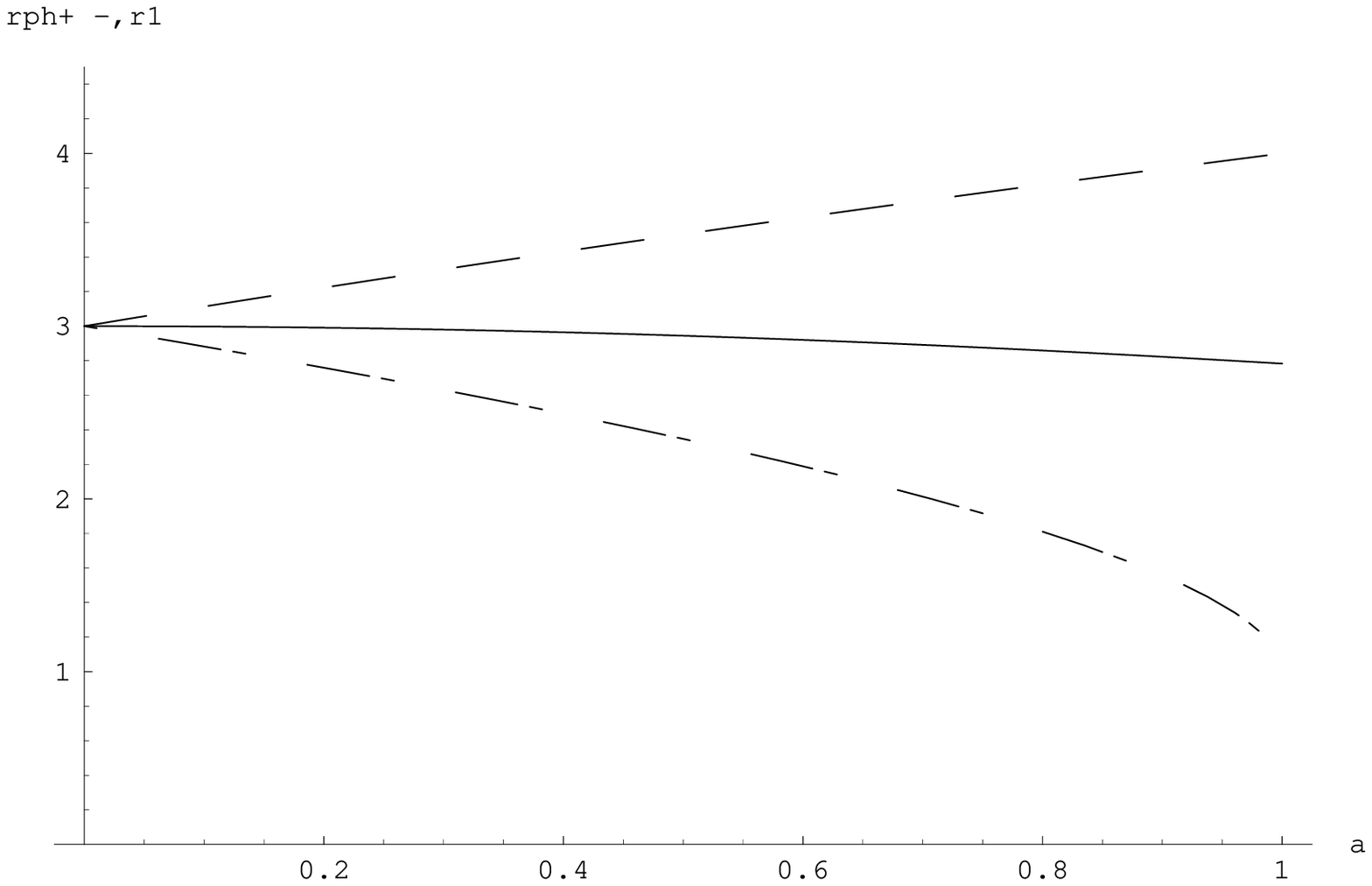}
\end{center}

\vspace*{-1.3in} \singlespace  Fig. 1. Location of the point where
$Cfg = 0$, (--------), retrograde photon orbit  (- - - -) and of
the prograde photon orbit (-$\cdot$ -$\cdot$ -$\cdot$) for
different values of
$a$.\\

\middlespace It may be easily seen that for $0 < a < 1$ the
equation (32) at best can have only three real roots of which, one
is always definitely outside the event horizon (Iyer and Prasanna
(1993)) as depicted in Fig. 1. However, as also shown in this
figure this location does not coincide with the location of the
unstable photon orbit, prograde or retrograde.  The centrifugal
force vanishes at a location between the two photon orbits and for
the case $a=0$, they all coincide at $r=3m$. Unfortunately, the
direct link between the unstable photon orbit and the centrifugal
force reversal, depicted in static spacetimes do not find a
parallel in stationary spacetime. Rotation does indeed bring in
some new
features of which the frame dragging is the most important one.\\

\middlespace Figs. 2 and 3 show the nature of the centrifugal and
Coriolis forces at the location of retrograde and direct photon
orbits as a function of $\Omega$ for $a = 0.5$. The first
impression that one gets is that these forces change sign for
different values of $\Omega$ across the asymptotes. However, one
has to check that the asymptotes are caused by the infinity of the
redshift factor $A^2$ at the roots of the equation
\be
\Omega^2 g_{\phi\phi} + 2\Omega g_{t\phi} + g_{tt} = 0 \ee

\be
\Omega_\pm = \o \pm \sqrt{\o^2-g_{tt}/g_{\phi\phi}} \ee

Hence the only portion of the plots which is meaningful is the
region corresponding to the values of $\Omega$; $\Omega_- < \Omega
< \Omega_+$. As may be seen in this region the centrifugal is
positive along the retrograde photon orbit ($rph+$) ane negative
along the direct photon orbit ($rph-$) as it should be according
to Fig. 1.  On the other hand, the coriolis changes sign in both
cases at the point $\Omega = \o$, wherein centrifugal also
vanishes because of the fact that the angular velocity is just
equal to the dragging of inertial frames, by the spacetime due to
the rotation of the central object.\\

\middlespace
 On the other hand, we have seen that there are
locations in the given spacetime, wherein the centrifugal force is
zero but $\Omega \neq \o$ and thus the Coriolis is non-zero.  In
view of this, Prasanna (1997) defined an index of reference called
the `Cumulative Drag Index' as defined by the ratio
\[
C = \f{\l( Co - Gr \r)}{\l( Co + Gr \r)}
\]
which could characterise purely the rotational feature of a
spacetime through its influence on a particle in circular orbit
at the location where the centrifugal force is zero.\\

\begin{center}
\epsfxsize 4in \epsfysize 4in \epsfbox{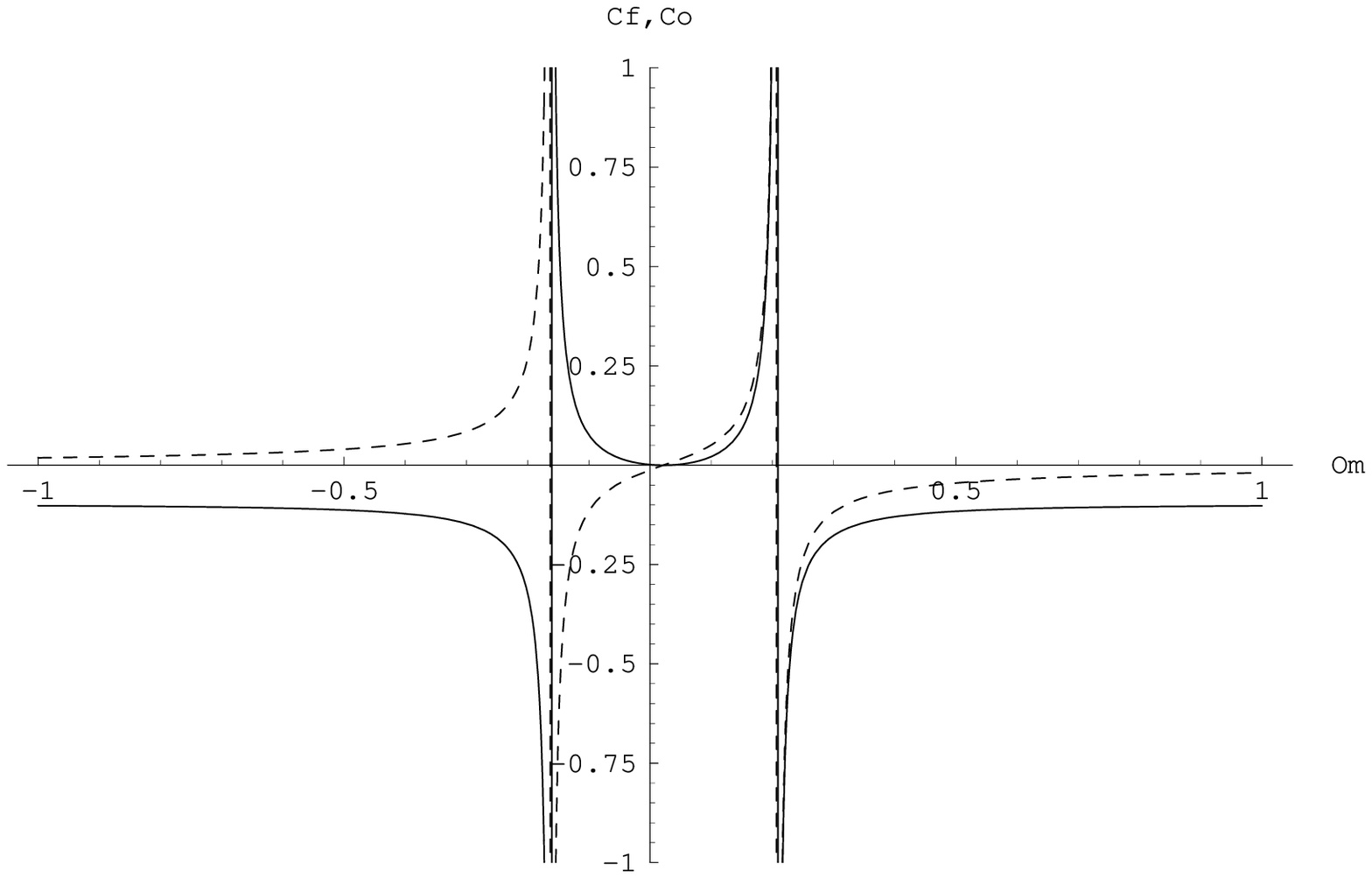}
\end{center}

\vspace*{-1in} \singlespace Fig. 2. Centrifugal (-------) and
Coriolis (- - -) force plots for $a = 0.5$, along the retrograde
photon orbit for $-1 < \Omega < 1$.\\

\begin{center}
\epsfxsize 4in \epsfysize 4in \epsfbox{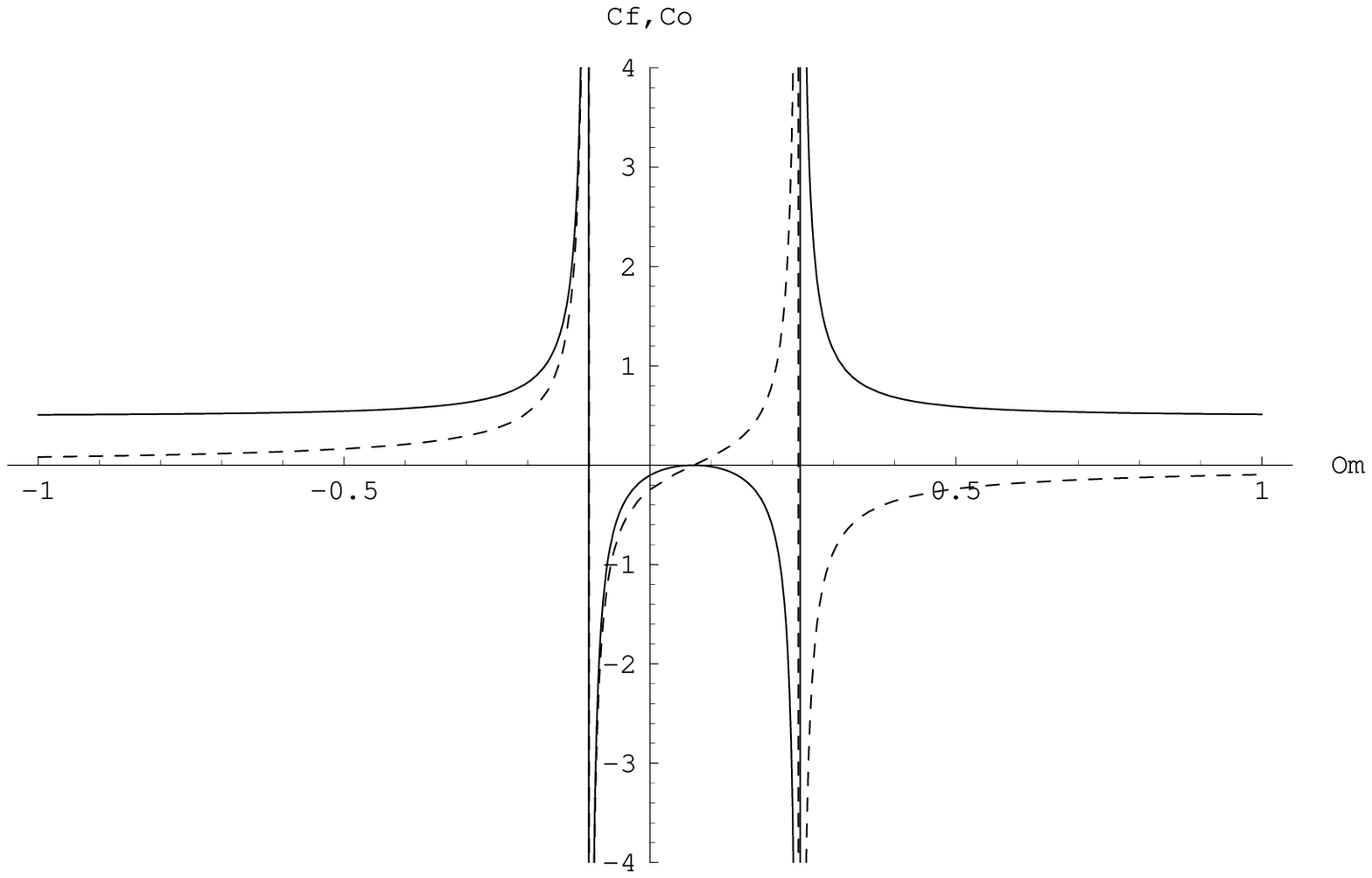}
\end{center}

\vspace*{-1in} \singlespace \centerline{Fig. 3. Same as (2) at the
prograde photon orbit.}

\newpage
\middlespace Fig. 4 shows the plot of $C$ as a function of
$\Omega$, for a fixed $a$ and $R$. As may be seen, there are two
zeros and two infinities for the function. As $a > 0$, $\Omega >
0$ represents the co-rotating particles and $\Omega < 0$
represents the
counter-rotating particles.\\

\begin{center}
\epsfxsize 4in \epsfysize 4in \epsfbox{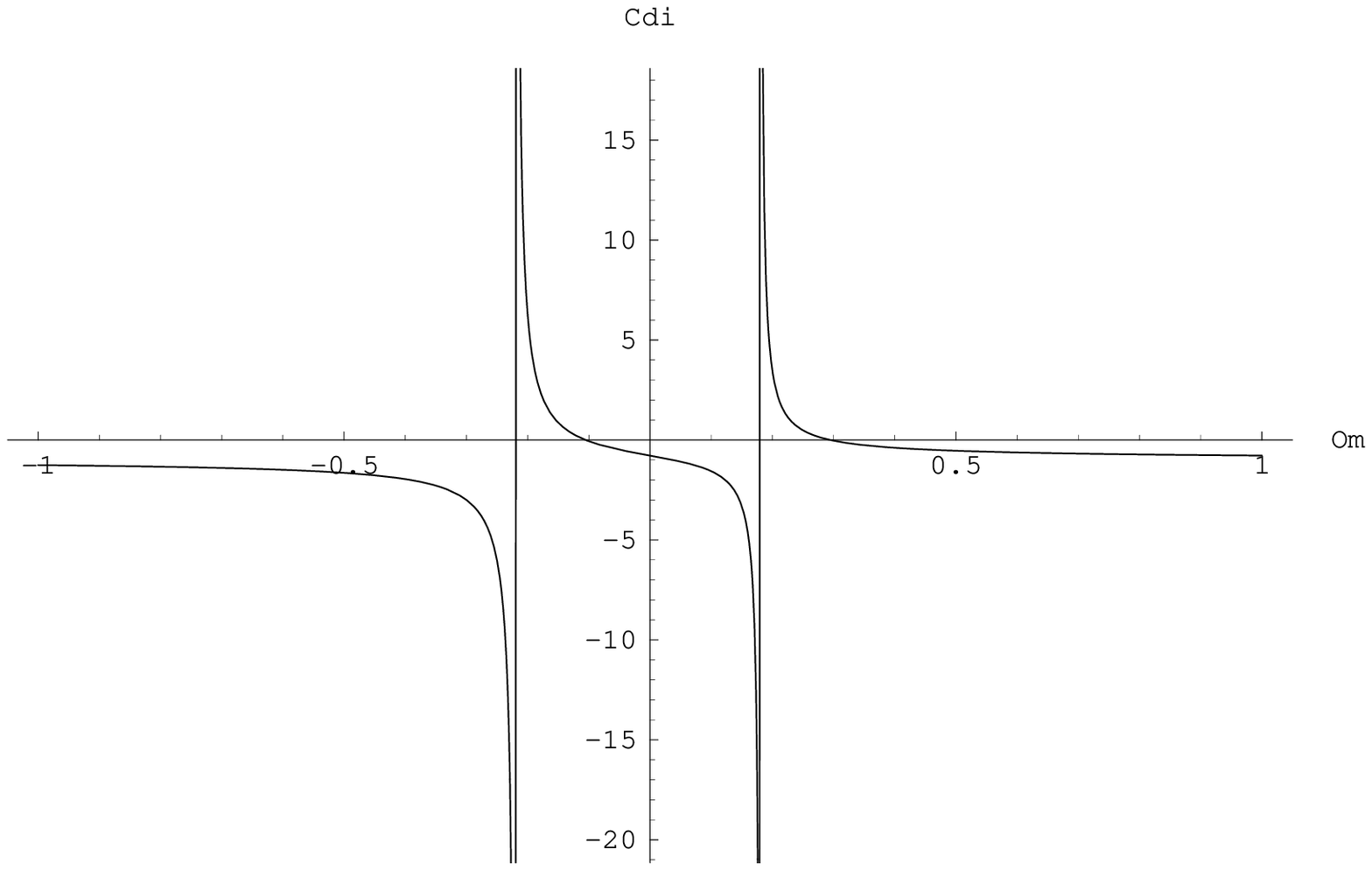}
\end{center}

\vspace*{-1.3in} \singlespace Fig. 4. Cumulative drag index $C$
for $a$ = 0.5, $R$ = 2.9445 as a function of $\Omega$ $(-1 <
\Omega < 1)$.\\

\middlespace

As the orbit chosen is the one where the centrifugal force is
identically zero the infinities of $C$ refer to the trajectories
along which the total force acting on the particle is zero as
given by the definition of $C$.  Fig. 5 shows the index $C$ for
three distinct cases of $a$ = 0.1, 0.5
and 1 at the respective locations where $Cf = 0$.\\

It is interesting to note that as $a$ increases, the co-rotating
particles have to decrease their angular velocity $\Omega$ very
little to stay in equilibrium, whereas the counter-rotating ones
have to increase their $\Omega$ much more, as depicted
explicitly in Table 1.\\
\newpage

\begin{center}
\epsfxsize 4in \epsfysize 4in \epsfbox{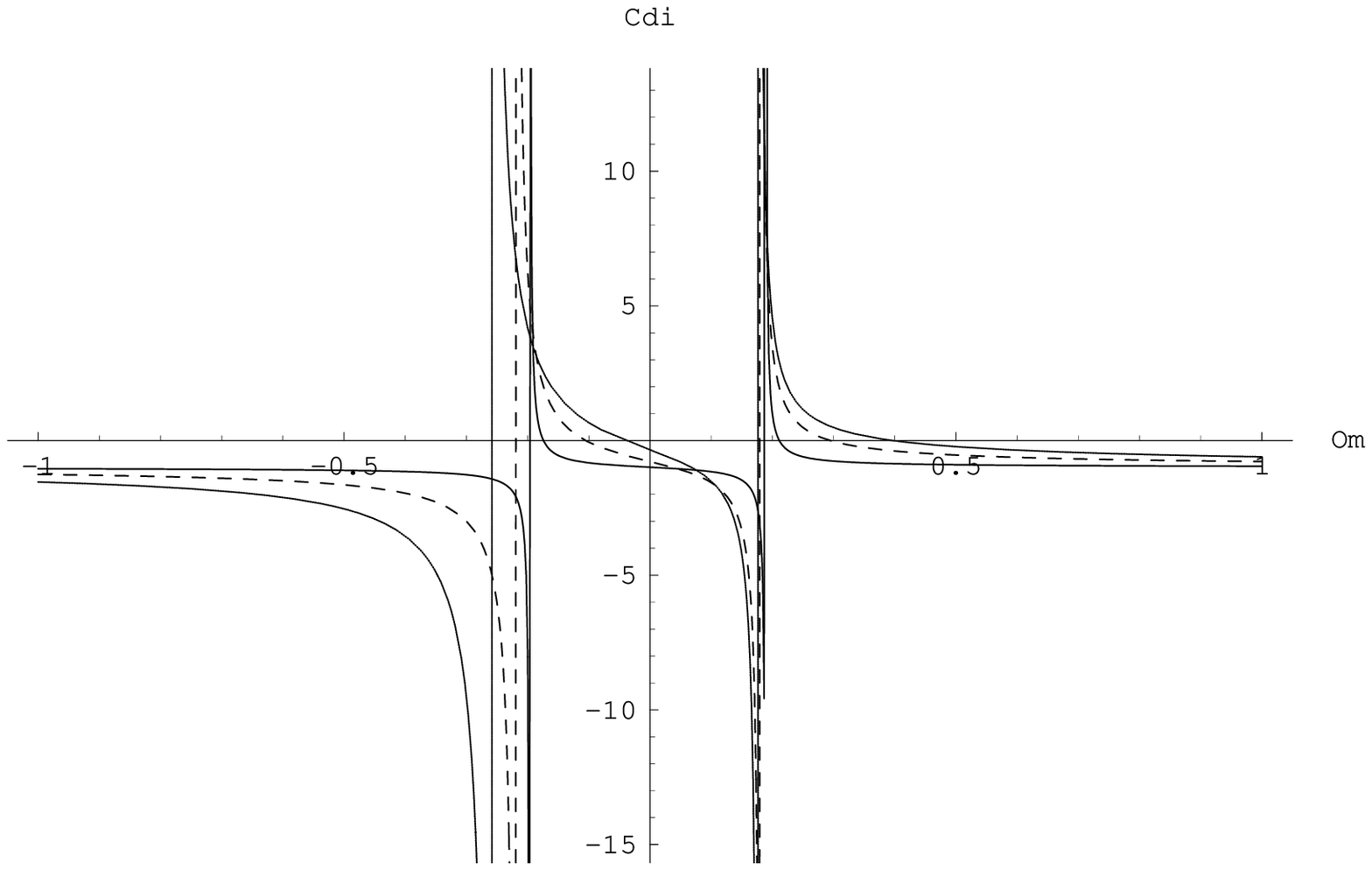}
\end{center}

\vspace*{-1.5in} \singlespace Fig. 5. $C$ for three different
values of $a$ showing difference in $\Omega_-$ values for
equilibrium orbit $\l( Co = - Gr \r)$.\\

\middlespace

\begin{center}
\underline{Table 1}\\[8pt]
\end{center}
\begin{center}
\begin{tabular}{||c|c|c|c|c|c||}
\hline\hline
$a$&$R$&$\Omega_+$&$\Omega_-$&$\left| \l( \Omega_- - \Omega_+
\r) \right|$&$\o$\\
\hline
0.1&2.9978&0.189022&-0.1964&0.0074&0.0074\\
\hline
0.5&2.9445&0.180094&-0.21965&0.0395&0.0374\\
\hline
1.0&2.7830&0.17722&-0.27452&0.097&0.076\\
\hline\hline
\end{tabular}
\end{center}

However, it may be seen that, the excess adjustment on the part
of counter-rotating particles is essentially due to the dragging
of inertial frames $\o$ which is always in the direction of
rotation of the central source. Table 1 gives the values of $\o$
for given $a$ and $R$, which matches quite closely with the
difference in the corresponding $\Omega$ for given $a$ and $R$,
between co- and counter-rotating particles. Thus one finds that
using the language of forces within the framework of general
relatvity, one can exactly quantify the spacetime curvature
effects which otherwise is hidden in the geometry.\\

There have been other approaches to describe particle motion in
general relativity a general survey of which was provided by
Bini et al. (1997). However, the conformal slicing as described
above helps directly in understanding certain physical features
of fluid flow configurations near extremaly compact objects and
also visualise the inertial drag more clearly.  It would indeed
be very nice if one can use this in the context of ADM formalism
for Hamiltonian dynamics.  This could be of direct consequence
to understand the dynamical collapse of fluid configurations,
through numerical approach.  It is indeed very necessary to
follow the collapse, particularly to see whether the centrifugal
reversal that is inherent would produce the strange behaviour of
the ellipticity as evidenced in the work of Chandrasekhar and
Miller.  A sudden change in ellipticity could indeed produce
noticeable signature on the emission of gravitational radiation
from a collapsing star in the final stages before it gets into a
blackhole.  This is an open problem which could be of great
interest for classical gravitation studies.\\

\newpage
\begin{center}
{\Large{\bf References}}\\
\end{center}

\begin{enumerate}
\item Arnowitt Deser S. and Misner C.W. (1962), in {\it
Gravitation - An Introduction to Current Research}, Ed. L.
Witten, p. 227-265.
\item Misner C.V., Thorne K.S. and Wheeler J.A. (1973) {\it
Gravitation}, Freeman \& Co.
\item Siedel E. (1996) {\it Numerical Relativity and Blackhole
Collision in Relativity and Scientific Computing}, Ed. F. Hehl
(Springer Verlag).
\item York J. (1979), in {\it Sources of Gravitational
Radiation}, Ed. L. Smarr, p. 83.
\item Abramowicz M.A., Carter B. and Lasota J.P. (1988) {\it G.R.G.
Journal},  {\bf 20}, p. 1173.
\item Abramowicz M.A., Nurowski and Wex (1993) {\it
Class. Quant. Gr.} {\bf 10}, L183.
\item Abramowicz M.A., Nurowski and Wex (1995) {\it
Class. Quant. Gr.}, {\bf 12}, p.~1467.
\item Bardeen J.M., Press W.H. and Teukolsky S.A. (1972)
{\it Astrophys. J.}, {\bf 178}, p. 347.
\item Prasanna A.R. (1997) {\it Class. Quan. Grav.}, {\bf 14}, p.
227.
\item Abramowicz M.A. and Prasanna A.R. (1990) {\it M.N.R.A.S.},
{\bf 245}, p. 720.
\item Anderson M.R. and Lemos J.P.S. (1988) {\it M.N.R.A.S.},
{\bf 233}, p. 489.
\item Chandrasekhar S. and Miller J.C. (1974) {\it M.N.R.A.S.},
{\bf 167}, p. 63.
\item Abramowicz M.A. and Miller J.C. (1990) {\it M.N.R.A.S.},
{\bf 245}, p. 729.
\item Gupta A., Iyer S. and Prasanna A.R. (1996) {\it Class.
Quan. Grav.}, {\bf 13}, p. 2675.
\item Prasanna A.R. (1991) {\it Phys. Rev. D} {\bf 43}, p. 1418.
\item Iyer S. and Prasanna A.R. (1993) {\it Class. Quan. Grav.},
{\bf 10}, L13.
\item Bini D., Carini P. and Jantzen R.T. (1997) {\it The
Inertial Forces - Test Particle Motion Game} (preprint).
\end{enumerate}

\end{document}